\documentclass[aps, prxlife, reprint, superscriptaddress, longbibliography]{revtex4-2}

\usepackage{graphicx}   
\usepackage{dcolumn}    
\usepackage{bm}         
\usepackage{amsmath,amssymb, bm}
\usepackage{mathtools}
\usepackage[inkscapelatex = false]{svg}
\usepackage{mleftright}
\usepackage{physics}
\usepackage{hyperref}
\usepackage{textgreek}

\newcommand{\est}{\hat\theta}
\newcommand{\true}{\bar \theta}

\newcommand{\cest}{c(\est,T)}

\newcommand{\error}{\delta \theta}
\newcommand{\uncert}{\epsilon^2}
\newcommand{\heading}[1]{\section{#1}}
\newcommand{\dlogP}{ \frac{d \log{P(\theta | \{t_i\})}}{d\theta}}
\newcommand{\sdlogP}{ \frac{d^2 \log{P(\theta | \{t_i\})}}{d\theta^2} }
\newcommand{\dc}[2]{\frac{\partial#1}{\partial #2}}

\newcommand{\atTheta}[1]{\Bigg\vert _{#1}}
\newcommand{\cinf}{c_{\infty}}

\begin{document}

\title{You ain't seen nothing, and yet:\texorpdfstring{\\}{}Future biochemical concentrations can be predicted with surprisingly high accuracy}

\author{Ketevan Danelia}
\affiliation{Department of Physics, Emory University, Atlanta, Georgia 30322, USA}

\author{Sean A.\ Ridout}
\affiliation{Department of Physics and Initiative for Theory and Modeling of Living Systems, Emory University, Atlanta, Georgia 30322, USA}
\author{Ilya Nemenman}
\affiliation{Department of Physics, Department of Biology, and Initiative for Theory and Modeling of Living Systems, Emory University, Atlanta, Georgia 30322, USA }
\date{\today}

\begin{abstract}
Accurate sensing of chemical concentrations is essential for numerous biological processes. The accuracy of this sensing, for small numbers of molecules, is limited by shot noise. Corresponding theoretical limits on sensing precision, as a function of sensing duration, have been well-studied in the context of quasi-static and randomly fluctuating concentrations. However, during development and in many other cases, concentration profiles are not random but exhibit predictable spatiotemporal patterns. We propose that leveraging prior knowledge of these structured profiles can improve and accelerate concentration sensing by utilizing information from current molecular binding events to predict future concentrations. By framing the constrained sensing problem as Bayesian inference over an allowed class of spatiotemporal profiles, we derive new theoretical limits on sensing accuracy. Our analysis reveals that maximum a posteriori (MAP) estimation can outperform the classical Berg-Purcell and maximum-likelihood (Poisson counting) limits, achieving a sensing precision of  $\delta c/c = 1/\sqrt{a^2N}$, where $N$ is the number of binding events, and $a > 1$ in certain cases. Thus knowledge of the statistical structure of concentration profiles enhances sensing precision, providing a potential explanation for the rapid yet highly accurate cell fate decisions observed during development.
\end{abstract}

\maketitle
\heading{Introduction} 
Understanding information processing in living organisms is essential for understanding life~\cite{tkavcik2016information, de2021self, levchenko2014cellular}. At the cellular level, information is often carried by spatiotemporally varying concentrations of signaling molecules~\cite{driever1988gradient,berridge1988spatial,niethammer2009tissue,atay2017spatial}. Thus elucidating the limits on the precision of sensing or estimation of molecular concentration set by the physics of molecular diffusion, molecule-receptor binding, and activation of downstream signals is a storied field of research. 

The field traces back to the work of Berg and Purcell~\cite{berg1977physics}, who derived the fundamental limits on concentration sensing by a sensor (e.g., a receptor) imposed by the physics of diffusion. While typically expressed differently, their results showed that the limit on the sensing uncertainty is $\delta  c/\bar c = 2/\sqrt{N(T)}$, where $N(T)$ is the total number of new ligand molecules that have been detected by the sensor during some observation time $T$. 
Notice that $1/\sqrt{N}$ is the na\"ive, ``Poisson'' benchmark for the uncertainty arising from the fluctuations in the number of binding events; the factor of 2 arises from the physics of the receptor binding, and from a bound receptor not providing any information about the measured concentration \cite{berg1977physics, endres2009maximum}. Subsequent studies incorporated various biologically relevant details, showing that more realistic features of the problem often increase the sensing error.  For example, ligand–receptor binding kinetics and rapid rebinding events generate temporal correlations that limit the number of independent measurements available to a cell, thereby degrading concentration estimation precision~\cite{bialek2005physical, kaizu2014berg}. Extensions to spatial sensing showed that gradient detection is fundamentally constrained by diffusive noise and receptor correlations across the cell surface, leading to poorer performance than na\"ive independent-receptor estimates~\cite{endres2009accuracy,hu2010physical}. Further, sensing over a large domain requires communicating the sensed information into one place, which is itself noisy and degrades performance \cite{ellison2016cell,mugler2016limits}. Similar limitations arise in sensing time-varying concentrations, where diffusion and receptor kinetics bound the accuracy with which temporal changes can be inferred~\cite{mora2010limits, mora2019physical}. Furthermore, correlations between receptors, arising from shared ligands or signaling pathways, further reduce information transmission compared to the Berg–Purcell idealization~\cite{singh2016effects}. 

The Berg--Purcell analysis has also been reformulated using inference- and information-theoretic approaches, showing that optimal decision strategies~\cite{siggia2013decisions, desponds2020mechanism} can improve sensing performance without changing the underlying physical scaling. Bayesian formulations have likewise clarified how prior information, geometry, and internal biases shape concentration and gradient sensing accuracy~\cite{hu2011geometry, novak2021bayesian}. Maximum likelihood estimation from receptor binding trajectories achieves lower variance (error) than simple time-averaged occupancy, while retaining Berg--Purcell-type scaling with the number of binding events~\cite{endres2009maximum}. Optimal filtering and decoding of time-dependent signals further improve estimation accuracy by exploiting temporal correlations, yet remain constrained by diffusion-imposed limits~\cite{malaguti2021theory,sartori2011noise}. At the network and multicellular levels, spatial averaging, receptor crosstalk, and collective signaling can enhance performance relative to isolated receptors, but remain fundamentally limited by molecular noise~\cite{smith2016role, mugler2016limits, singh2017simple,carballo2019receptor}. Optimized signaling architectures and spatial profiles can buffer noise and improve robustness of downstream readout without violating fundamental physical limits~\cite{paulsson2004summing,saunders2009morphogen,tostevin2009mutual}.  Collectively, these results significantly contributed to our understanding of the limits of concentration sensing accuracy across various contexts. 

Yet all of these results focused on either stationary concentrations or temporally randomly varying concentrations. In contrast, real biological systems often lie between these extremes: concentration profiles are neither static nor purely stochastic but are expected to follow relatively  predictable spatiotemporal patterns~\cite{struhl1989gradient, rink1989calcium, niethammer2009tissue}. A salient example is organismal development, where cell‑fate decisions depend on positional information encoded by morphogen gradients~\cite{wolpert1969positional, driever1988gradient, smith2009forming, gregor2007stability, rogers2011morphogen, wieschaus2016positional}. These gradients are certainly not stationary and not purely random; they are results of a complex, dynamic interplay of many molecular players, and they vary little from one embryo to another within a species~\cite{gregor2007stability}. From the perspective of a sensor, knowing that the sensed concentration can only come from a restricted set of patterns that are possible in this particular developing organism is a strong {\em a priori} constraint on the sensing  problem. In principle, such prior information can be used in a Bayesian fashion to reduce the estimation variance, as has been explored previously in gradient-sensing settings~\cite{hu2011geometry, novak2021bayesian}. In development, where morphogen concentrations often reach steady states at long times, having such {\em a priori} information may even allow the sensing cell to {\em predict} the eventual steady state concentration, long before it is reached. Thus, while one typically assumes cellular fate commitment based on steady state concentration values, perhaps a cell does not need to wait for the steady state to know what it will be and to commit.   This raises a fundamental question: can living organisms exploit such constrained  structure of concentration profiles to predict future concentrations and/or surpass classical sensing limits? 

To answer this question, here we first formulate the problem of concentration sensing accuracy when a prior over allowable concentration profiles is available by casting it as a maximum a posteriori (MAP) Bayesian estimation, instead of a more traditional maximum likelihood (ML) problem \cite{endres2009maximum}. Inspired by embryonic development, and by minimal models that reproduce some features of established morphogen gradients ~\cite{crick1970diffusion, driever1988gradient, gregor2007stability, kicheva2012investigating} we focus on monotonically saturating concentration profiles. Under this assumption, we derive the concentration sensing limits analytically and confirm them with numerical simulations. We show that, while the Poisson-like Berg-Purcell scaling $\delta c/c\propto 1/\sqrt{N}$ is preserved, the prefactor depends on the  details of the concentration profiles. Thus, better-than-Poisson estimation accuracy is possible for concentration profiles where the maximum rate of change is strongly correlated with the eventual steady state concentration value. 

As an example to illustrate our framework, we consider a simplified production--diffusion--degradation model of Bicoid gradient formation in \emph{Drosophila} development. We show that this model produces concentration profiles that lie precisely in the regime where our formalism predicts better-than-Poisson estimation accuracy. Thus, we argue that this super-Poisson sensing can contribute to solving a long-standing puzzle of developmental precision in {\em Drosophila}, although this simple picture neglects complications like cytoplasmic flows and nuclear motion, which should be treated in future work. Finally, we show that the required Bayesian computations can be implemented by relatively simple biochemical networks, suggesting that cells may be able to utilize such sensing (and prediction!) strategies.

\heading{Concentration Estimation with Bayesian prior}
We consider a sensor device (e.g., a receptor) that is exposed to a concentration profile $c(t)$. The profile is sampled from an underlying probability distribution indexed by a parameter $\theta$, $P[c(t)|\theta]$. The parameter is unknown to the device; thus the concentration is unknown as well. The parameter $\theta$ can correspond to the position of the receptor in the embryo, or to the (future) saturated value of the concentration, or it can be some other index of the concentration profile. For generality, we do not specify what $\theta$ is until it becomes necessary. 

We assume that the receptor has access to the molecular  binding sequence \( \{t_i\} \), with \( i = 1, \dots, N \), over the time interval \([0, T]\), and that the binding events are probabilistic readouts of the underlying concentration. The goal is to estimate the parameter $\theta$. This can be done using Bayes' theorem:  
\begin{equation}
\label{eq: Bayes Formula}
    P(\theta | \{t_i\}) = \frac{P( \{t_i  \} | \theta )P(\theta)}{ P( \{t_i\} ) }.
\end{equation}
Here, $ P(\{t_i\} | \theta) $ represents the likelihood of observing the binding sequence $ \{t_i\} $ given $\theta$,  and $P(\theta)$  denotes the prior distribution of $\theta$, which we assume is also known to the sensor.  In other words, the inference problem is not to estimate an arbitrary function, but to choose which member of a known family generated the data. Since the binding sequence depends on concentration and not on $\theta$ directly,  the likelihood $ P(\{t_i\} | \theta) $ can be expressed as a functional integral over all configurations of the concentration field $c_\theta(t)$ {\em a priori} allowed in the problem:
\begin{align}
\label{eq: funct integral expanded likelihood}
    P( \{t_i  \} | \theta ) &= \int P(\left \{t_i  \} |c_\theta( t)  \right) P( c_\theta( t) | \theta ) \mathcal{D}[c_\theta( t)],
\end{align}
where, $ \mathcal{D}[c_\theta(t)]$ denotes integration over all allowable concentrations, and $ P(\{t_i\} | c_\theta(t))$ is the likelihood of observing the binding times $ \{t_i\} $ given a specific concentration value at time $t$, $c_\theta(t)$. 
For simplicity, we assume  instantaneous and independent bindings with a diffusion-limited binding rate $r=4DA$, where $A$ is the linear size of the sensor. 
We do not model the receptor bound times and unbinding---these simply renormalize the time the receptor is available for sensing \cite{endres2009maximum,singh2017simple,mora2019physical}. Then $ P(\{t_i\} | c_\theta(t))$ is a given by a time-dependent Poisson process:
\begin{equation}
\label{eq: binding/unbinding prob}
    P(\{t_i\} | c_\theta(t)) = e^{- \int_0^T dt\,   r c_\theta(t)  }\prod_{i=1}^N r c_\theta( t_i).
\end{equation}
Here, the exponential term $ e^{ - \int_0^T dt\,   r c_\theta(t) } $ accounts for the probability of  no events occurring between the observed binding times, and the product $ \prod_{i=1}^N r c_\theta( t_i) $ describes the probability of binding at each  binding time $ t_i $.

The second factor in Eq.~\eqref{eq: funct integral expanded likelihood}, $P(c_\theta(t) |\theta)$, is the prior probability density of the concentration profile $c_\theta(t)$ given $\theta$. While, in reality, even knowing the parameter $\theta$ does not deterministically specify $c(t)$ due to the usual stochasticity of production, degradation, and diffusion of the sensed molecules, here we further simplify the problem and consider the relationship between $\theta$ and $c_\theta(t)$ deterministic. Thus, 
\begin{equation}
    P(c_\theta(t) |\theta)=\delta(c_\theta(t) - \bar{c}_{\theta}(t)|\theta), 
    \label{eq: delta-pdf}
\end{equation}
where $\bar{c}_{\theta}(t)$ is the deterministic concentration profile given $\theta$. 
In other words, {\em all}   stochasticity in the problem comes only from the probabilistic receptor bindings. We extend these results to the case of small stochastic fluctuations of the concentration around $\bar{c}_\theta(t)$ in Appendix A. Such fluctuations do not change the MAP estimator at lowest order, but lead to a controlled renormalization of the estimation error.

Substituting Eqs.~(\ref{eq: binding/unbinding prob}, \ref{eq: delta-pdf}) into Eq.~\eqref{eq: funct integral expanded likelihood}, and performing the integral over the $\delta$-distribution, gives: 
\begin{align}
\label{eq: Posterior distribution}
    P(\theta | \{t_i\}) = \frac{1}{Z}P(\theta)  e^{- \int_0^T dt   r \bar{c}_{\theta}(t) } \prod_{i=1}^N \bar{c}_\theta(t_i). 
\end{align}
Here $Z = {r^N}/{P(\{t_i\})}$, where $r^N$ is a constant absorbed into the normalization factor $1/{P(\{t_i\}})$.

Now we can find the MAP estimate $\hat\theta$ of the parameter by maximizing the log of the posterior 
\begin{equation}
\label{eq: Map est general}
    \dlogP\atTheta{\est}  =  0.
\end{equation}
To approximate the variance of this estimator, $\epsilon^2$, we evaluate the curvature of the log-posterior around the MAP estimate. 
In the large-$N$ limit, where the posterior becomes sharply peaked, the variance of an unbiased estimator can be approximated by the inverse of the Fisher information, as given by the Cram\'er–Rao bound~\cite{kay1993fundamentals}
\begin{equation}
      \label{eq: relation between second derivative and variance}
    \uncert = -\frac{1}{\left\langle\sdlogP\right\rangle \atTheta{\bar\theta}},
\end{equation}
where $\bar \theta$  is the true value of a parameter.
For our particular model, Eq.~(\ref{eq: Map est general}) gives:
\begin{multline}
    \label{eq: first derivative of log posterior}
    \int_0^T \left(\frac{\partial \log \bar{c}_{\theta}(t)}{\partial \theta}\sum_i^N \delta(t-t_i) -r\dc{\bar{c}_{\theta}(t)}{\theta}\right)dt  \atTheta{\est} \\= -\frac{d \log{P(\theta)}}{d\theta}\atTheta{\hat \theta}.
\end{multline}
This equation must be solved at a fixed $T$ to obtain the estimate $\hat{\theta}(T)$ at that time.  
Solving  Eq.~\eqref{eq: first derivative of log posterior} analytically for an arbitrary concentration profile $c_\theta(t)$ is impossible. However, we can explore a few special cases and approximations to understand the behavior of the solution.

First, as $T\to\infty$, we expect any MAP estimator to concentrate near the true value. Thus, 
\begin{equation}
\label{eq: ML estimation}
  \est (T)= \true +  \error(T),
\end{equation}
where $\delta\theta$ is  the sensing error with the (small) variance of $\epsilon^2(T)$. 
To determine this variance $\uncert(T)$,  we simplify  Eq.~\eqref{eq: relation between second derivative and variance}. Specifically, we compute the second derivative of the log-likelihood  with respect to the parameter, average over the ensemble of all possible binding event sequences, and evaluate the result at the true parameter $\bar\theta$. 
The stochastic sum over binding events,  $\sum_1^N \delta (t - t_i)$, simplifies under ensemble averaging to  $r\bar{c}_{\bar \theta}$, and we finally have:
\begin{equation}
\label{eq: final variance}
    \uncert (T)= \frac{1}{-\frac{\partial^2\log{P(\theta)}}{\partial\theta^2} \atTheta{\bar\theta}+\int_0^T \left.\frac{r}{\bar{c}_{\theta}}\left(\dc{\bar{c}_{\theta}}{\theta}\right)^2  \right|_{\bar{\theta}}dt }.
\end{equation}

To the extent that the integral in the denominator scales as the average number of binding events, $\epsilon^2\propto1/N(T)$. 
However, while Eq.~\eqref{eq: final variance} naively suggests  $\epsilon^2\propto 1/T$ as well, if $\bar{c}$ has an explicit time dependence (as we would expect in various development problems), the scaling of $\epsilon^2$ with $T$ may be different. 
Note also that the exact prefactor in front of $1/N$ would depend on  how sensitively the concentration profile $\bar{c}_{\theta}(t)$ depends on the parameter $\theta$. In other words, binding events  carry information directly only about the {\em concentration}, and the accuracy of the  estimation of the parameter $\theta$ depends on how one needs to transform the concentration estimate to obtain the parameter from it. 

To place these results in context, consider now a simple case of a concentration profile independent of $t$,  $\bar{c}_\theta(t)=\bar{c}_\theta(t\to\infty)\equiv c_\infty$.  
Defining in this case $\theta=c_\infty$ (that is, the concentration is parameterized by its value), we transform Eq.~\eqref{eq: first derivative of log posterior} into an equation for the MAP estimate $\hat{c}_\infty (T)$ at time $T$:  
\begin{equation}
    \int_0^T dt \left( \frac{1}{\hat{c}_\infty(T)}\sum_i^N \delta(t-t_i) -r\right) = -\frac{d\log{P(\hat c_\infty)}}{d\hat c_\infty}.
\end{equation}
Performing the integration, and dropping the $O(1/\sqrt{N})$ contribution from the prior, immediately gives: 
\begin{equation}
    \label{eq:c_const}
\hat{c}_\infty \simeq \frac{N(T)}{rT} .
\end{equation}
Similarly, Eq.~\eqref{eq: final variance} becomes:
\begin{equation}
\label{eq:var_const}
    \uncert (T)\simeq \frac{c_\infty}{rT}, \quad
\frac{\uncert}{\cinf^2} \simeq \frac{1}{N(T)},
\end{equation}
where, in all cases, $a \simeq b$ denotes equality of $a$ and $b$ at leading-order in $1/N$. 

In other words, in the constant concentration case, the MAP estimate reduced to the  simple estimate we expect for a Poisson binding sequence---just dividing the number of binding events by the elapsed time, with a dimensional normalization by the binding rate.  This result recovers the classical limit of ML estimation under constant concentration, consistent with previous results~\cite{endres2009maximum}. As mentioned above, this differs by the factor of 2 from  the Berg-Purcell result \cite{berg1977physics}, because the latter symmetrically utilizes both bound and unbound receptor states, despite bound times carrying no concentration information, whereas ML estimation optimally weights only the informative unbound intervals~\cite{endres2009maximum}.

For a more general $\bar{c}_\theta(t)$ at   $T\to\infty$, we note that, if $T$ increases by a small amount $dT$, then only a few new bindings will happen, and thus the estimate $\hat{\theta}$ will change only by a small amount $d\hat{\theta}$ as well. 
Thus, to get some intuition for the general behavior of the $\hat{\theta}$, we can transform Eq.~\eqref{eq: first derivative of log posterior} into a (stochastic) differential equation describing the dynamics of $\hat{\theta}$ as a function of the measurement time $T$. For this, we define $F(\hat\theta, T)$: 
\begin{multline}
    \label{eq: estimate evolution in time}F(\est, T) \equiv  
      \frac{d \log{P(\theta)}}{d\theta}\atTheta{\hat \theta}+\\\int_0^T \left(\frac{\partial \log \bar{c}_\theta(t)}{\partial \theta}\sum_i^N \delta(t-t_i) -r\dc{\bar{c}_\theta(t)}{\theta}\right)dt  \atTheta{\est} \\= 0.
      \end{multline}
Now, notice that, by definition, $F$ does not change  over a small time interval, so that    $F(\hat{\theta}+\Delta \hat{\theta}, T+\Delta T) = 0$. Thus expanding  in small $\Delta T, \Delta\hat{\theta}$ gives: 
\begin{equation}
    \dc{F(\est,T)}{\est}\Delta \hat{\theta}  + \dc{F(\est,T)}{T} \Delta T +O(\Delta T^2,\Delta\hat{\theta}^2,\Delta T\Delta\hat{\theta}) = 0.
\end{equation}
As we take $\Delta T \to 0$, the $O(\Delta T^2)$ terms vanish. Yet, since $\hat{\theta}$ jumps at each binding event, the $O(\Delta \hat{\theta}^2)$ terms do not necessarily vanish.  However such jumps are expected to be $\lesssim\epsilon(N)$, since an individual binding event should not affect the estimator by more than its current uncertainty. Thus, at large $T$, when $N(T)\gg1$, and $\epsilon(N)\to0$, a discrete jump in $\hat{\theta}$ due to an individual binding can still be viewed as small, resulting in 
\begin{equation}
           \label{eq: meth of char}
      \dc{F(\est,T)}{\est}\frac{d\est}{dT} + \dc{F(\est,T)}{T}=0.
\end{equation}
By isolating $d\est/dT$ from Eq.~\eqref{eq: meth of char}, dropping the contribution from the prior $P(\theta)$, which does not scale with $T$ or $N$ and therefore should not matter in the large $N$ limit, and noting that $\partial F(\est,T)/\partial\est= \sdlogP\atTheta{\est}= -\frac{1}{\uncert}$, we arrive at:
\begin{gather}
\label{eq: estimate dynamics}
    \frac{d\est}{dT} \simeq \uncert(T) \left( \frac{\partial \log \bar{c}_{\theta}(T)}{\partial \theta} \atTheta{\est}\sum_i \delta(T - t_i) -r\dc{\bar{c}_{\theta}(T)}{\theta}\atTheta{\est}\right).
\end{gather}
In this large-$N$ regime, since the effect of the prior vanishes, the estimator effectively reduces to an ML estimator (we return to this point in Section \ref{sec:beating}); in other regimes, however, the prior may play an important role~\cite{hu2011geometry} . We stress that this equation is only valid for $N\gg1$, where a jump in $\hat{\theta}$ at each binding event is small and the contribution of the prior becomes negligible. Note also that this is, indeed, a stochastic ODE since binding times $t_i$ are random.

Suppose now that there are no binding events $t_i$ in the immediate vicinity of $T$. Then the first term in Eq.~\eqref{eq: estimate dynamics} is zero, and
\begin{gather}
\label{eq:no_binding}
    \frac{d\est}{dT} = -r\uncert(T)   \dc{\bar{c}_{\theta}(T)}{\theta}\atTheta{\est}.
\end{gather}
Thus, if the parameter is the concentration itself, so that the partial derivative in Eq.~\eqref{eq:no_binding} is unity, then, in the absence of bindings, the estimate of the concentration falls with time, weighted by $\epsilon^2$ and $r$. In other words, one expects $r c_{\hat{\theta}}$ bindings per unit time, so that the absence of binding events is an indication that the concentration is overestimated, and the estimate must decrease. However, this new evidence must be weighed against all the prior evidence encapsulated in $\epsilon^2$. And if $\dc{\cest}{\est}\neq 1$, then the change in the concentration estimate must be reparameterized into the parameter estimate.

If $T$ is near some $t_i$ in Eq.~\eqref{eq: estimate dynamics}, we can integrate the equation over a tiny range, between $t_{i-}$ and $t_{i+}$, to obtain the change in $\hat{\theta}$ across a binding event, $\Delta\hat{\theta}$. The contribution of the second term in the rhs of Eq.~\eqref{eq: estimate dynamics} vanishes, the $\delta$-function integrates to one, and we get:
\begin{gather}
\label{eq:jump}
    \Delta\hat{\theta} \approx { \uncert(t_i) } \frac{\partial \log \bar{c}_{\theta} (t_i)}{\partial \theta} \atTheta{\est}.
\end{gather}
In other words, the concentration estimate receives a bump proportional to $\epsilon^2$, and the change in the estimate of a general parameter $\theta$ is then the reparameterized change in the concentration estimate. Note again that, as discussed above, Eq.~\eqref{eq:jump}  is an approximation valid when the update $\Delta \hat{\theta}$ is small; the exact $\Delta \hat{\theta}$ must solve a nonlinear equation.

Overall, Eq.~\eqref{eq: estimate dynamics} paints the following picture: the concentration estimate falls between bindings and receives a bump at a binding, as in \cite{mora2019physical}, with new binding events becoming progressively less informative and less important when $\epsilon^2$ decreases with $T$. Thus the concentration estimator undergoes a stochastic relaxation in a potential with an ever-increasing stiffness. The rightward bumps of the estimator are driven by stochastic binding events, and the leftwards relaxation is deterministic. With time, the relaxation and the stochastic bindings balance each other, and the estimate settles near a true concentration value. Further,  if the problem is parameterized by a general parameter $\theta$ instead of the concentration, then it can be obtained from the concentration estimate via a simple reparameterization.

\begin{figure}
    \centering
    \includegraphics[width =\columnwidth]{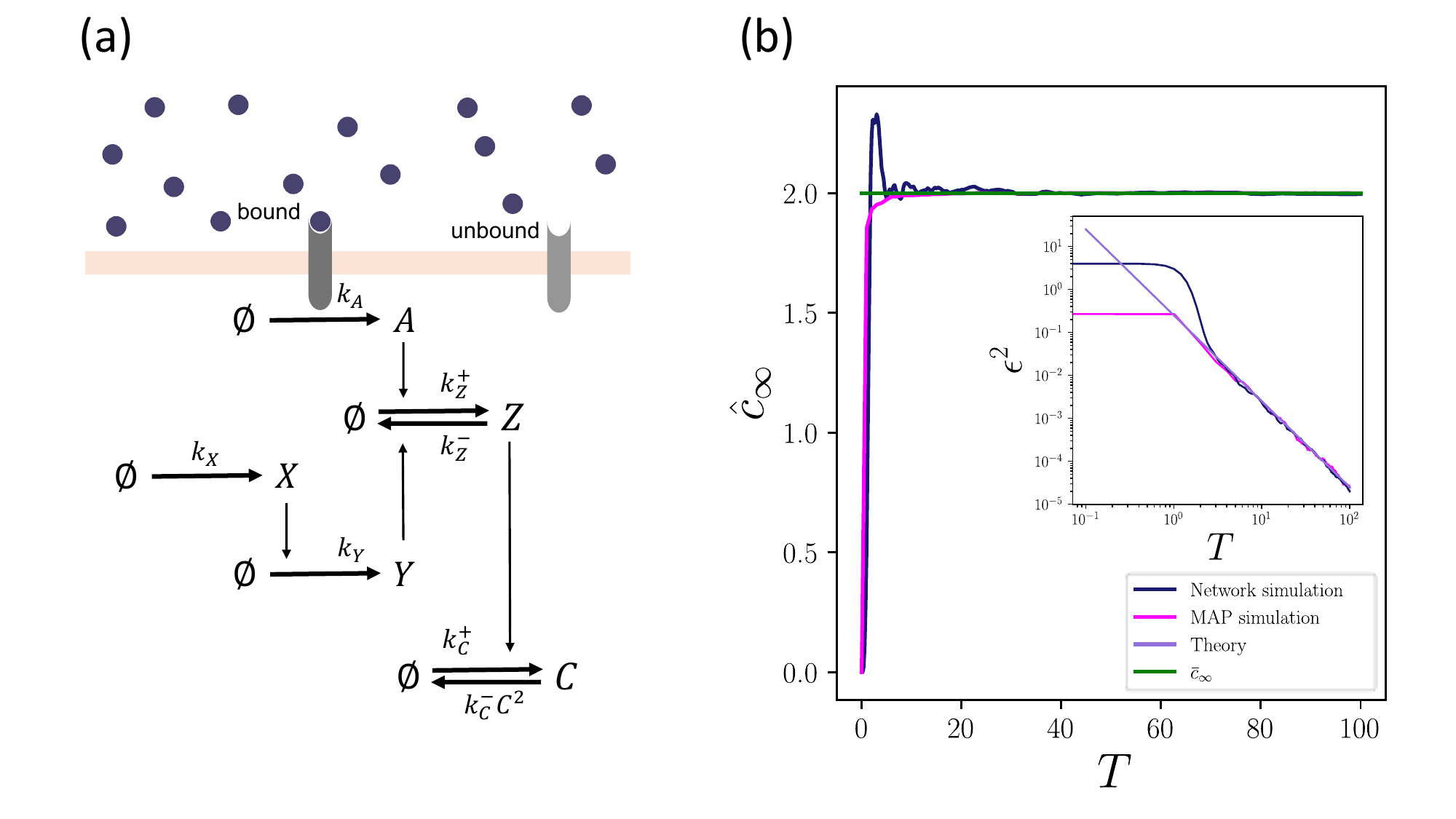}
    \caption{Concentration estimation by a  biochemical network. (a) Schematic of the biochemical network performing Bayesian inference. Receptor binding event activates the readout molecule $A$. (b) Simulation of the network readout in response to stochastic binding events in a linearly increasing concentration field compared with the exact MAP estimate (without approximations),  the approximate theoretical MAP estimate, and their corresponding errors. }
    \label{fig:biochemical netwok}
\end{figure}

To verify calculations in this section, we generated binding event sequences $\{t_i\}$ from an inhomogeneous Poisson process with time-dependent rate $\lambda(t)=r c_\theta(t)$ and $c_\theta(t)=k\theta^a t$. 
For these concentration profiles, because of the explicit time dependence of the concentration profile,  Eq.~\eqref{eq: final variance} reduces to $\uncert(T) = 2/a^2krT^2$. Indeed, numerical averages over $n=100$ repetitions produced  estimator means and variances consistent with our calculations here in the $N\gg1$ limit (see Fig.~\ref{fig:biochemical netwok}(b)) .  
\begin{figure*}{}
    \centering
    \includegraphics[width = 7.06 in]{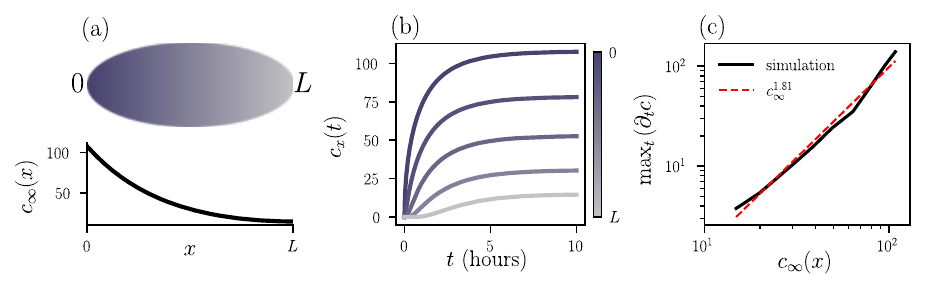}
    \caption{(a) The elliptical shape represents a {\em Drosophila} embryo during early development of size $L\approx 500  \textrm{\,\textmu m}$. 
    The grayscale gradient illustrates the steady state concentration profile of the Bicoid protein along the anterior--posterior axis, which is established within approximately one hour after fertilization. Panel (b) shows simulated spatiotemporal evolution of Bicoid gradient, for various positions $x$. 
    Simulations use a minimal diffusion-degradation model for Bicoid gradient formation~\cite{driever1988gradient} with constant production $j$ at the anterior end, diffusion coefficient $D$,  linear degradation with rate $r$, and reflecting boundary condition at the posterior end. This is the simplest model reproducing many features of the  experimentally observed concentration profile, although more complex models are needed for better quantitative accuracy~\cite{gregor2007stability, bergmann2007pre, abu2010high}. Parameter values are $(D, j , r, L) \simeq  (5 \textrm{\,\textmu m}^2 \mathrm{s}^{-1},  5 \textrm{\,\textmu m}^{-2}\mathrm{h}^{-1}, 0.5\textrm{\,h}^{-1}, 500 \textrm{\,\textmu m})$ and the units of concentration are $\textrm{\textmu m}^{-3}$. Panel (c) shows that, in this model,  the peak temporal rate of change of the concentration, $\max \partial_t c_x(t)$, scales as a $\sim 1.8$ power of the steady-state level $\cinf(x)$ at the same spatial position. }
    \label{fig:drosophilla example}
\end{figure*}
\heading{Beating the Poisson Counting Limit} \label{sec:beating} Consider Bicoid gradient sensing during {\em Drosophila} early development, see Fig.~\ref{fig:drosophilla example}(a). The goal of a sensor (a nucleus in the embryo) at position $x$ is to estimate its position from the sequence of binding events~\cite{wieschaus2016positional}. If one waits for an hour or two post-fertilization, an approximately exponential stationary concentration profile is established~\cite{abu2010high}, Fig.~\ref{fig:drosophilla example}(a, b). One can then estimate the stationary concentration with the Poisson accuracy, as in Eqs.~(\ref{eq:c_const}, \ref{eq:var_const}), by counting the binding events in the steady state. Then transforming from $c_\infty$ to $x$ is just a reparameterization.
 \begin{figure*}
    \centering
    \includegraphics[width = 7.06 in]{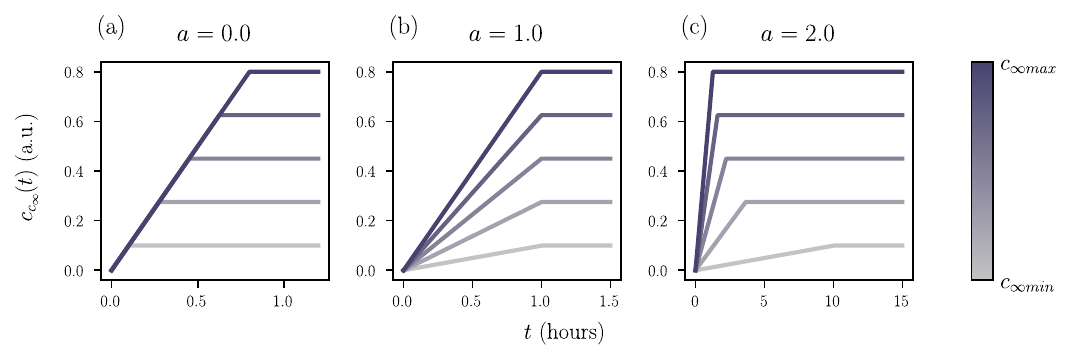}
    \caption{Model piecewise-linear concentration profiles defined in Eq.~\eqref{eq: piecewise profile with a} for different values of the parameter $a$. The color of each curve corresponds to its final saturation value $\cinf$.  }
    \label{fig:piecewise profiles}
\end{figure*}

Thus, we now consider a  broader class of \textit{time-varying} profiles,  where the concentration evolves as a scaled version of its eventual saturation value $c(t) = \cinf f(t)$,  with $f(t\to\infty) = 1$. 
Here, $f(t)$ is an arbitrary monotonic function describing the temporal evolution, while $\cinf$ parametrizes the final steady-state concentration. This is equivalent to (very crudely) approximating the concentration profiles in Fig.~\ref{fig:drosophilla example}(b) as reaching position-specific saturation at a uniform time over the whole embryo. Then
Eqs.~(\ref{eq: first derivative of log posterior}, \ref{eq: final variance}) transform into:
\begin{gather}
    \label{eq:c f(t)}
    \int_0^T dt \left( \frac{1}{\hat{c}_\infty(T)}\sum_i \delta(t-t_i) -r f(t)\right) = -\frac{d \log{P(c_\infty)}}{dc_\infty}\atTheta{\hat c_\infty},\\
    \label{eq: var for cf(t)}
    \uncert(T) = \frac{1}{-\frac{\partial^2\log{P(c_\infty)}}{\partial c_\infty^2} \atTheta{\bar c_\infty} +\int_0^T \frac{rf(t)}{\bar{c}_\infty} dt} .
\end{gather}
We perform the integrals, noting that  $\int_0^T \frac{rf(t)}{\bar c_\infty}dt  = \int_0^T \frac{r\bar c_\infty f(t)}{\bar c_\infty^2}dt = \frac{\bar N(T)}{\bar c_\infty^2} $, where $\bar{N}(T)$ is the expected number of binding events given the true $\bar{c}_\infty$. 
Note  that $\bar N(T) = N(T) + O(\sqrt{N(T)})$, and thus, in the large $N$ limit, we write $\bar N(T) \simeq N(T)$. We also note that the prior terms are lower order in $N(T)$. Thus, we obtain
\begin{equation}
\hat{c}_\infty \simeq \frac{N(T)}{r\int_0^T f(t) dt},\quad
\frac{\uncert}{\bar{c}_\infty^2} \simeq \frac{1}{N(T)}, \label{eq:scaling_est_err}
\end{equation}
where again $\simeq$ denotes leading order for large $N, T$.  In the limit that these corrections can be ignored, our estimator becomes an ML estimator. Notice, however, that this is the ML estimator for the specific class of possible concentration profiles, not for a constant concentration as in \cite{endres2009maximum, mora2015physical, singh2017simple}

Note that the sensing error is still Poisson, as in Eq.~(\ref{eq:var_const}), but it goes down with {\em total} number of binding events and not the number since reaching the steady state.
This suggests that, in this case, the time dependence of the concentration does not change the fundamental Poisson scaling, but it does allow the eventual saturation value of the concentration, and hence the sensor position, to be estimated (predicted!) with Poisson accuracy long before that concentration is reached.

Figure \ref{fig:drosophilla example} (c) shows that this approximation, in which the concentration evolves as a scaled version of its eventual saturation value, is too crude. Even within our simple one-dimensional diffusion--degradation model of the Bicoid gradient, we show in Fig.~\ref{fig:drosophilla example} that the concentration profile cannot be written as a position-dependent rescaling of a universal time-dependent curve, and thus Eq.~\eqref{eq:scaling_est_err} does not apply. Indeed, the maximum rate of change of the concentration $v_c(x)=\max_t \partial_t c_x(t)$  is a nonlinear function of $c_\infty(x)$: higher saturation values mean faster approach to the steady state, $v(x)\sim c_\infty^{1.8}$ in this simple model.  
Different parameter values (degradation rate and the diffusion constant) result in different, sometimes even stronger, dependence between $c_\infty(x)$ and $v_c(x)$. 

We will now develop a simplified, analytically-tractable model to show the effects of this nonlinear scaling. Firstly, note that $c_x(t)$ in Fig.~\ref{fig:drosophilla example}(b) has three phases: nearly zero before the diffusing front arrives, rapid growth, and saturation. The initial phase, during which no binding events have yet occurred, may itself contain information about $c_\infty$: cells closer to the source generally wait less for the first binding event. Using this information, however, would require the cell to know when development started, necessitating a more complex biochemical implementation. For simplicity, we thus assume that the cell begins its estimation when the first molecule of signal is detected, shifting $t=0$ for each position to mark the time the concentration begins to rise sharply, skipping the initial delay (in Section \ref{sec:biochem}, we will briefly discuss how this could be implemented biochemically). Our analysis will thus produce an \emph{upper bound} on the achievable sensing error, which could be improved even further by an algorithm that extracts information from the initial delay. 

To analyze inference that begins with the first binding event, we model the growth and saturation phases of the profile. We approximate the  dependence in Fig.~\ref{fig:drosophilla example}(b,c) with a piecewise linear concentration profile, Fig.~\ref{fig:piecewise profiles}:
\begin{equation} 
\label{eq: piecewise profile with a} 
c(\cinf, t) =  
\begin{cases}
k \cinf^{a} t , & t < \cinf^{1-a}/k \equiv t_1, \\
\cinf, & t \geq t_1,
\end{cases}
\end{equation}
where, again $c_\infty=c_\infty(x)$ is the $t\to\infty$ saturating value of the concentration at the position $x$. Here, $k$ has the units of $[\text{concentration}]^{1-a}/\text{s}$. 
For $a = 1$, this functional form is a special case of the scaled profiles discussed above.  Note that this  piecewise-linear profile is not a crucial part of the analysis; it merely provides an analytically tractable representative of a broad class of monotone saturating profiles, where the maximum rate of change and the saturating concentration are strongly coupled. 

For this piecewise-linear concentration, assuming measurement happens during some time interval $T > t_1$, our MAP estimator takes the form:
\begin{equation}
    \label{eq: est for entire piecewise profile}
    \hat{c}_\infty \simeq \ \frac{a N_1 + N_2}{rT +r t_1 (a/2 -1)}, 
\end{equation}
where again $\simeq$ stands for equality to the leading order in $N,T$. Here, $N_1 \text{ and } N_2$ denote the numbers of binding events that occurred during the concentration growth phase ($t \le t_1$) and after saturation  ($t_1 < t \leq T$), respectively. 
Thus binding events before and after saturation are weighed differently by the estimator. Importantly, the estimator depends on $t_1$, which itself is a function of the saturation value $c_\infty$, and would have to be estimated self-consistently. Implementing such an estimator biochemically would require retaining memory not only of how many binding events occurred, but also of {\em when} they occurred. We will discuss how this problem might be solved biochemically near the end of Section \ref{sec:biochem}.

The corresponding estimation error is:
\begin{equation}
    \label{eq: error for wntire piecewise}
    \frac{\uncert}{\cinf^2} \simeq \frac{1}{a^2 N_1 +N_2}.
\end{equation}
If we assume that measurement time $T\leq t_1= \cinf^{1-a}/k$, so that all measurements are still in the concentration growth phase, the MAP estimator and its error are:
\begin{equation}
\hat{c}_\infty \simeq \left(\frac{2N(T)}{rkT^2}\right)^{1/a},\quad
\frac{\uncert}{\cinf^2} \simeq \frac{1}{a^2 N}.
\label{eq:superpoisson}
\end{equation}
The dependence on $T^2$ in the first of these equations is easy to understand---from Eq.~\eqref{eq: piecewise profile with a}, the expected number of binding events over time $T<t_1$ is $\bar N(T)=rkc^a_\infty T^2/2$, which immediately results in $\hat{c}_\infty=c_\infty$.

The second of Eqs.~\eqref{eq:superpoisson} says that, as long as $a>1$, so that there is a strong positive correlation between the early rate of change of the concentration and the eventual saturation, the uncertainty is {\em better than Poisson} by a factor  of $a^2$. For $a\sim1.8$ as in Fig.~\ref{fig:drosophilla example}(c), this corresponds to about a 70\% accuracy improvement--- quite a substantial improvement. 

Thus, with this MAP algorithm, the concentration (and hence the position in the embryo) can be \emph{predicted} before steady-state concentration is reached. The accuracy of this  prediction still scales as $1/N$, but our specific class of allowed profiles produces a prefactor that can be smaller than the standard Poisson value. More generally, this \emph{super-Poisson} accuracy will be seen whenever the expected concentration profiles are designed in a way so that subtle differences in $c_\infty$  manifest in large changes in the statistics of binding events early in the estimation process.

Since morphogen concentrations are spatially structured and interpreted by nuclei to determine their position and consequently their future fate, it is natural to ask how accurately position, and not just the eventual saturating concentration, can be inferred from local measurements.  We thus examine the position estimation accuracy under the classical ML model for constant concentrations versus our MAP framework.  We take the piece-wise linear saturating profile with   $c_\infty=c_\infty(x)$ in Eqs.~(\ref{eq: final variance}, \ref{eq: piecewise profile with a}). The actual dependence $c_\infty(x)$ can be a power-law or an exponential, with the details  not important for the derivation below, as long as the dependence is smooth and monotonic.  Under these conditions, Eq.~(\ref{eq: final variance}) becomes: 
\begin{equation}
    \uncert_x = \frac{c_\infty^2(x)}{a^2 \left(\frac{dc_\infty(x)}{dx}\right)^2{N(T)}},
\end{equation} 
where $\uncert_x$ is the uncertainty of the position estimator, and, again, $N = rkc_\infty^a T^2/2$. Recall that the classical constant-concentration estimator has $\uncert_x = \left| \frac{dx}{dc_\infty} \right| ^2 \frac{c^2_\infty(x)}{N(T)}$. Thus, the same profile-dependent function of $x$ converts between the concentration error and the positional error in both the MAP and the ML estimators, and the improvement in positional inference in our model mirrors the improvement in concentration prediction.

\heading{Plausible Biochemical Networks for Performing MAP Estimation} \label{sec:biochem}
Can biochemical networks perform the computations required for MAP estimation? For arbitrary expected spatiotemporal concentration profiles, these computations could be very complex, so that no general solution exists. Thus constructing a network capable of performing MAP estimation requires specifying the profile class, as well as making additional assumptions about the involved  parameter values and dynamical regime (pre- or post-saturation), in which the network is expected to operate. However, our case is simpler: when all terms in Eq.~\eqref{eq: estimate dynamics} are rational functions of time, implementing the necessary computations in biological wetware should be possible. As a proof of principle, we construct a network that solves the problem for the pre-saturation regime $T < k\cinf^{1-a}$ for the piecewise linear profiles (Eq.~\ref{eq: piecewise profile with a}) with $a=2$. 

In this case ($a=2$, pre-saturation), Eq.~\eqref{eq: estimate dynamics} yields:
\begin{equation}
    \label{eq: cinf evolution in time}
    \frac{d \hat c_\infty}{dT} = \frac{1}{rkT^2\hat c_\infty} \sum_i \delta(T-t_i) -\frac{\hat c_\infty}{T}.
\end{equation}
For a biochemical network to implement this computation, it only needs an estimate of time, $T$ and $T^2$, since the beginning of the process, in  addition to the variable  estimating $c_\infty$ {\em per se}, and then to take ratios of these quantities. Luckily, counting time can be done with constitutive production, squaring or taking products of concentrations is just a homo-dimer or a hetero-dimer formation, and inversion of a concentration $x\to1/x$ can be approximated, for example, by $x$ being a deactivating enzyme for some substrate. In other words, every algebraic operation in Eq.~\eqref{eq: cinf evolution in time} can be performed biochemically.

For the concentration profile in Eq.~\eqref{eq: piecewise profile with a}, one can design an even simpler biochemical computer which calculates the MAP estimator given in Eq.~\eqref{eq:superpoisson}. A pair of molecular species $X, Y$ are used to track time and its integral, while another species $A$ is produced every time a binding event of the measured molecule is detected. If $A$ acts as a catalyst for the production of a readout species $Z$, and $Y$ a catalyst for its degradation, then this allows the number of binding events to be normalized by the time squared in quasi-steady state (ss): 
\begin{align}
\label{eq: Biochem net} 
\frac{dA}{dT}=k_A \sum_i \delta(T-t_i)\quad&\Rightarrow\quad A=k_AN(T),\\
\frac{dX}{dT}=k_X\quad&\Rightarrow\quad X=k_XT, \label{eq:X_prod}\\
\frac{dY}{dT}=k_Y X\quad&\Rightarrow\quad Y=\frac{k_Xk_Y}{2}T^2,\\
\frac{dZ}{dT}=k_Z^+A - k_Z^-YZ\quad&\xRightarrow{\mbox{\small ss}}\; Z_{\mathrm{ss}}=\frac{2k_Z^+k_A}{k_Z^-k_Xk_Y}\frac{N}{T^2}.\label{eq:last}
\end{align}
Thus, as long as $\frac{k_Z^+k_A}{k_Z^-k_Xk_Y}=\frac{1}{rk}$, $Z_{\mathrm{ss}}$ is the square of $\hat{c}_\infty$. In principle, one can infer positional information from $Z_{\mathrm{ss}}$ directly. However, if one really wants to have an explicit readout of  $\hat{c}_\infty$ from $Z_{\mathrm{ss}}$, taking a square root is again easy in steady state,  by introducing a final readout species $C$~\cite{mora2019physical}: 
\begin{equation}
\frac{dC}{dT}=k_C^+Z-k_C^-C^2\;\xRightarrow{\mbox{\small ss}}\; C_{\mathrm{ss}}= \sqrt{\frac{2k_Z^+k_Ak_C^+}{k_Z^-k_Xk_Yk_C^-}\frac{N}{T^2}},  
\end{equation}
which matches $\hat{c}_\infty$ in Eq.~\eqref{eq:superpoisson} with a suitable choice of rate constants.

To test whether this biochemical reaction network can implement optimal inference, we simulate stochastic receptor binding events and compare an explicit MAP estimator of the saturation concentration $c_\infty$ to a dynamical biochemical realization of the same computation. Binding events are generated as an inhomogeneous Poisson process with time-dependent rate $\lambda(t) = rk c_\infty^at$ corresponding to the pre-saturation dynamics with, for concreteness,  $\bar c_\infty = 2$, and $a=2$. The MAP estimate is computed by evaluating the log-likelihood on a discrete grid of candidate $c_\infty$ values using only binding events up to time $T$. Tracking this estimate as $T$ increases allows us to follow both its convergence and its mean-squared error, which are then averaged over $n =100$ realizations. 
The same binding sequences quantify the performance of the  biochemical network, Eqs.~(\ref{eq: Biochem net}--\ref{eq:last}). The agreement between the biochemical network output and the MAP estimator, both in the mean and in the scaling of the error with time, shows that nearly-optimal inference can be realized by compact biochemical reaction networks driven by stochastic binding events. Note that Fig.~\ref{fig:biochemical netwok}(b) also shows agreement between both explicit MAP estimator and its variance converging to our theoretical results derived in Eqs.~(\ref{eq: ML estimation}, \ref{eq: final variance}).

The network constructed above assumed linear profiles in the pre-saturation regime with $a=2$. Relaxing these assumptions would require constructing a network tailored to other regimes. In particular, designing a network for non-integer values of $a$ is not straightforward, as it would require a biochemical implementation of non-integer powers of the running estimate (see~Eq.~\eqref{eq: estimate dynamics}). Similarly, once the signal enters the saturating regime, the present network is no longer expected to remain unbiased without additional circuitry that detects this transition and modifies the update rule accordingly. However, such an extension is straightforward in principle, and we now sketch an algorithm for how it can be implemented.

Recall that the MAP estimator for the full profile requires classifying binding events as occurring before or after the saturation time $t_1(c_\infty)$ ($N_1$ and $N_2$ in Eq.~\ref{eq: est for entire piecewise profile}), since the two regimes are weighted differently. The exact implementation therefore requires full memory of the binding times to \emph{reclassify} binding events as the estimate $t_1(\hat{c}_\infty)$ is updated.
This exact calculation, however, is unnecessary. Including the rising phase of $c(t)$ is only worthwhile if $t_1 = O(T)$, where $T$ is the final time of the inference. If $t_1 \ll T$, then the information gained during the rising phase is negligible compared to that accumulated in steady state, and the simpler estimator based on $N_2$ alone would suffice. Assuming $t_1 = O(T)$, by the time the network needs to switch from accumulating $N_1$  to $N_2$, the saturation time $t_1$ has already been estimated with the error of $O(t_1/\sqrt{N_1})$. 
Thus, we expect that the MAP estimator can be adequately approximated by a network that ``switches'' from classifying events into $N_1$ to classifying them into $N_2$ at a single time, without needing to reclassify old events. This estimator requires no detailed memory. In the particular case of $a=2$, this ``switching'' is easy to implement: all profiles have $\bar{N}(t_1)=r/(2k)$, and thus the sensor should switch its behavior when it detects $A>k_A r/(2k)$. This switching could be implemented by a bistable feedback loop~\cite{alon2019introduction}, switching states irreversibly and sequestering the existing $A$ into a stable complex, thus routing new binding events into a new counter $N_2$.

Finally, recall that the biochemical network in Eqs.~(\ref{eq: Biochem net}--\ref{eq:last}) carries out MAP inference for piecewise linear profiles that begin to rise at $t=0$. However, the profiles in Fig.~\ref{fig:drosophilla example}(b) also had an initial delay. Incorporating this delay does not require any change in the dynamics of $A$ (Eq.~\ref{eq: Biochem net}) because no or very few binding events will occur before $c$ begins to rise. To produce the correct estimator, however, we require that $X$ and $Y$ reflect the time since the concentration began to rise, rather than the time since development began. This can be achieved by making $X$ production conditional on $A$ exceeding a small threshold (modifying Eq.~\ref{eq:X_prod}), effectively starting the estimator at the first binding event. 

Fully working out these extensions, including biochemical implementations for non-integer $a$, circuitry for detecting the saturation, and for omitting the initial phase, is beyond the scope of the present work. We therefore regard the current construction not as a universal biochemical realization of MAP sensing, but as an existence proof in a particularly simple case. 

\heading{Discussion}
We developed a  Bayesian formalism for incorporating prior knowledge of spatiotemporal concentration profiles into biochemical concentration estimation. We showed, in particular, that if the ensemble of concentration profiles is known, one can {\em predict} the concentration at a certain time (e.g., at the steady state), rather than wait to estimate it. Thus, for example, a nucleus can estimate its position in the embryo without waiting for the morphogen gradient that carries this information to stabilize.

Interestingly, the achievable sensing error is determined not only by the number of binding events, but also by how strongly the early-time binding statistics depend on the parameter being inferred. In our piecewise linear model, the early signal scales as $c_\infty^a$ , so the differences in the eventual saturation values are amplified into larger differences in early-time observed data. This changes the prefactor of the $1/\sqrt{N}$ scaling, without changing the scaling itself. MAP inference is then the procedure that allows the cell to attain this improved accuracy by extracting $c_\infty$ from the unsaturated, time-dependent trajectory. The gain is the largest when the nonlinear dependence is strong---that is, for large $a$. It is intriguing to speculate if the morphogen gradients in development have evolved to increase this sensitivity and hence maximize the estimation accuracy.

In some of the  models we considered, the total number of sensed molecules, $N(T)$, was a sufficient statistic of the entire binding sequence, so that the estimated parameter depended only on the number of molecules detected, but not on when they were detected. Equation~(\ref{eq: est for entire piecewise profile}) generalized this, so that two statistics contributed: the number of molecules captured before and after the saturation. In yet more complicated spatiotemporal concentration profiles, especially those not monotonic in time, the temporal distribution of the binding events is likely to be even more important, providing additional information about the estimated parameter, and, potentially, improving the estimation accuracy even more.

While our results here have been developed for simplified minimal models, they are likely to affect  our quantitative understanding of  development. In {\em Drosophila} early development, experimental studies show persistent puzzles~\cite{gregor2007probing,tkavcik2008information, manu2009canalization, garcia2013quantitative}: for example, nearby nuclei after division $14$ reliably exhibit distinct gene expression patterns, and the expression is initiated $\sim 3$ minutes after the division. This is $\sim 40$ times faster than the classical Berg-Purcell limit \cite{berg1977physics}  for how long one needs to sense to estimate the local concentration of a morphogen to a necessary precision~\cite{gregor2007probing}. Ways of resolving the contradiction have been proposed, such as ``on-the-fly'' sensing~\cite{siggia2013decisions,desponds2020mechanism}, but whether real biological systems utilize such approaches is still debated.  In our analysis, the estimation precision continues to scale as $1/N(T)$. Yet, with the molecular machinery adapted to sensing of specific developmental concentration profiles, this $N(T)$ can count molecules starting from fertilization, and not just from when concentrations  stabilized after the last division~\cite{ali2016bicoid, huang2017decoding}, which can increase the sensing period (and hence the sensing accuracy) many-fold. Crucially, for concentration profiles where concentrations rise faster near the origin of the morphogen, as in Fig.~\ref{fig:piecewise profiles}, the error prefactor can also be smaller than one.  Although we illustrated this effect with a very simple model of morphogen gradient formation, its \emph{sign} is likely to be extremely general: for any process where a signaling molecule moves outward from a source, we expect the concentration to rise sooner in the same places where the final concentration is higher. These two effects---not waiting for the steady state and the higher precision---can easily combine to improve the estimation quality by a factor of a few, contributing to resolving the precision puzzle.

Whether such  MAP estimation and prediction can be achieved in practice remains an open question. While simple integrations, multiplications, and divisions that are needed for the MAP solution for the profiles in Fig.~\ref{fig:piecewise profiles} are easily attainable, the concentration profiles experienced by sensors during embryogenesis are more complicated. In a fly, they depend not just on the morphogen gradients, as we modeled here, but also on how nuclei move in these gradients  between and during divisions ~\cite{gregor2007stability,hernandez2023two}. An exciting open problem is to characterize the  empirical probability distributions of these spatiotemporal profiles, derive MAP solutions for them, and to understand whether the network of regulatory interactions during early development can be understood not just as maximizing extraction of positional information in the steady state \cite{sokolowski2025deriving}, but as estimating the detector's position from the sensed {\em dynamical} concentration profiles.

A related limitation of our framework is the assumption that the parameters defining the trajectory class, such as $k$ and $a$, as well as the binding rate $r$, are known. This prior knowledge could be ``learned'' through evolution, and is encoded by the cell in the structure of the estimating biochemical network and in the concentrations of enzymes, through $k_X, k_Y$, and $k_Z^{\pm}$. This, however, is not special to our formalism: the classical Berg--Purcell bound likewise must treat quantities such as the diffusion coefficient, receptor size, and integration time as known, since the measured signal depends on them jointly with the concentration. Indeed, no matter what algorithm is used, there is no free lunch: exploiting temporal structure requires prior knowledge of that structure, and a biochemical network designed to implement MAP estimation for one class of profiles will produce a bias if used outside that class.  The simple network we presented is a proof of principle, and a different network would need to be used in, e.g., \emph{Drosophila}, to match the detailed physics of the morphogen gradient formation therein. Understanding how inference degrades under parameter variability, uncertainty, or partial non-identifiability remains an important direction for future work. 

Finally, beyond development, leveraging the spatiotemporal structure of signals is likely to occur in other processes that depend on tightly pre-programmed, reproducible concentration profiles, such as wound healing~\cite{niethammer2009tissue}. Making these connections and comparing the precision of these processes to those set by our theory is likely to lead to interesting insights. 

\begin{acknowledgments}
This work was supported, in part, by the NSF grant No.\ 2209996 to IN and by the Simons Foundation Investigator Award to IN. We thank Thierry Mora for stimulating discussions. 
\end{acknowledgments}

\bibliography{references} 
\onecolumngrid
\appendix

\section{Effect of small stochastic fluctuations around the mean concentration}

In the main text, we considered the case where, given the parameter $\theta$, the concentration trajectory was deterministic. However, in addition to stochastic binding events, the sensed concentration itself can be somewhat stochastic, even in a pre-programmed system, such as development. For example, the production of the morphogen is itself stochastic, and even the enzymes that produce it also fluctuate (intrinsic and extrinsic stochasticity in the terminology of \cite{elowitz2002stochastic}). Here we show how to treat this stochasticity in concentration sensing. 

Recall that the parameter estimation in our problem starts with the Bayes' rule:
\begin{equation}
P(\theta\mid \{t_i\}) = \frac{P(\{t_i\}\mid \theta)\, P(\theta)}{P(\{t_i\})}.
\tag{1}
\end{equation}
The likelihood can be expressed by marginalizing over all possible concentration trajectories $c(t)$,
\begin{equation}
P(\{t_i\}\mid \theta)
= \int \mathcal D c(t)\, P(\{t_i\}\mid c(t))\, P(c(t)\mid \theta),
\tag{2}
\end{equation}
or, equivalently, as an average over trajectories drawn from the prior $P(c(t)\mid\theta)$,
\begin{equation}
\label{eq: avg likelihood}
P(\{t_i\}\mid \theta)
= \big\langle P(\{t_i\}\mid c(t)) \big\rangle_{c(t)\sim P(c(t)\mid\theta)}.
\end{equation}

The conditional likelihood of observing a sequence of binding events $\{t_i\}$ given a concentration trajectory $c(t)$ is identical to that used in the main text and corresponds to an inhomogeneous Poisson process
\begin{equation}
\label{eq: binding/unbinding prob2}
P(\{t_i\}\mid c(t))
=
\exp\!\left(- r\int_0^T c(t)\,dt\right)
\prod_{i=1}^N r\,c(t_i).
\end{equation}
Substituting Eq.~\eqref{eq: binding/unbinding prob2} into Eq.~\eqref{eq: avg likelihood}, the posterior distribution becomes
\begin{equation}
\label{eq: post dist with avg signs}
P(\theta \mid \{t_i\})
=
\frac{1}{Z} P(\theta)
\left\langle
\exp\!\left(- r\int_0^T c(t)\,dt\right)
\prod_{i=1}^N \,c(t_i)
\right\rangle_{c(t)\sim P(c(t)\mid\theta)},
\end{equation}
where $Z = r^N/P(\{t_i\})$ is a normalization constant. 

The only difference between the present treatment and that in the main text is the choice of prior over concentration trajectories. Instead of a $\delta$--function prior, we now allow for small stochastic fluctuations around the mean trajectory, treating $c(t)$ as a narrow log-normal variable, so that it is always positive \cite{mora2019physical}.  Specifically, let $\delta y(t)$ be a Gaussian process with zero mean and covariance $K_\theta(t,t')$, and define 
\begin{equation}
c(t) = \bar c_\theta(t)\,
\exp\!\left[-\tfrac12 K_\theta(t,t) + \delta y(t)\right].
\end{equation}
The shift by $-\tfrac12 K_\theta(t,t)$ in the exponent ensures that $\langle c(t)\rangle = \bar c_\theta(t)$, so that $\bar c_\theta(t)$ retains its interpretation as the mean concentration profile indexed by $\theta$.

To obtain the maximum a posteriori (MAP) estimate and its uncertainty, we analyze the logarithm of the posterior distribution,
\begin{equation}
\label{eq:logmean}
\ell(\theta)
\equiv \log P(\theta\mid\{t_i\})
=
-\log Z + \log P(\theta)
+
\log
\left\langle
\exp\!\left(-r\int_0^T c(t)\,dt\right)
\prod_{i=1}^N c(t_i)
\right\rangle_{c(t)\sim P(c(t)\mid\theta)} .
\end{equation}
The first two terms can be neglected: $-\log Z$ is independent of $\theta$, and $\log P(\theta)$ does not scale with either the observation time $T$ or the number of binding events $N$, and is therefore negligible in the asymptotic regime of interest. We thus approximate
\begin{equation}
\label{eq: ell of theta simplified}
\ell(\theta)
\approx
\log
\left\langle
\exp\!\left(-r\int_0^T c(t)\,dt\right)
\prod_{i=1}^N c(t_i)
\right\rangle_{c(t)\sim P(c(t)\mid\theta)} .
\end{equation}
Introducing the functional
\begin{equation}
S_\theta[c] 
= 
\sum_{i=1}^{N} \log c(t_i)
-
r\int_0^T c(t)\,dt,
\end{equation}
we may write $\ell(\theta) \approx \log \langle e^{S_\theta}\rangle$. To proceed analytically, we assume that the fluctuations $\delta y(t)$ are small, so that the distribution of trajectories $c(t)$ is narrowly concentrated around the mean profile $\bar c_\theta(t)$. In this regime, the random variable $S_\theta = S_\theta[c]$ fluctuates weakly, motivating a cumulant expansion of $\log\langle e^{S_\theta}\rangle$. 

Retaining only the first two cumulants, we obtain
\begin{equation}
\ell(\theta)
\approx
\langle S_\theta\rangle
+
\frac12\,\mathrm{Var}(S_\theta).
\end{equation}
This approximation is controlled in the limit of small covariance $K_\theta$ and captures all corrections linear in the strength of the concentration fluctuations.
Explicitly, this yields
\begin{equation}
\label{eq: log cumulant 2}
\ell(\theta)
\approx
\Big\langle
\sum_{i=1}^{N} \log c(t_i)
-
r\int_0^T c(t)\,dt
\Big\rangle_{c(t)}
+
\frac12
\mathrm{Var}\!\left[
\sum_{i=1}^{N} \log c(t_i)
-
r\int_0^T c(t)\,dt
\right].
\end{equation}
To evaluate the cumulants appearing in Eq.~\eqref{eq: log cumulant 2} we substitute the explicit form $ c(t)=\bar c_\theta(t)\exp\!\left[-\tfrac12 K_\theta(t,t)+\delta y(t)\right],$ where $\delta y(t)$ is a zero-mean Gaussian process with covariance $K_\theta(t,t')$.  Terms with  $\log c(t)$ can be handled exactly, because
\begin{equation}
\log c(t)=\log\bar c_\theta(t)-\tfrac12 K_\theta(t,t)+\delta y(t).
\end{equation}
The first cumulant is obtained by averaging term by term. Using the fact that $\langle\delta y(t)\rangle=0$ and the mean concentration satisfies $\langle c(t)\rangle=\bar c_\theta(t)$ we get
\begin{equation}
\Big\langle \sum_{i=1}^N \log c(t_i) - r\int_0^T c(t)\,dt \Big\rangle_{c(t)}
=
\sum_{i=1}^N \left[\log\bar c_\theta(t_i)-\tfrac12 K_\theta(t_i,t_i)\right]
-
r\int_0^T \bar c_\theta(t)\,dt,
\end{equation}
To compute the second cumulant, we note that all fluctuations arise from the Gaussian field $\delta y(t)$. In the small noise regime, we want to calculate this cumulant to leading order in $K_\theta$. Thus, we linearize \begin{equation}
\label{eq: small-noise-linearization}
    c(t)=\bar c_\theta(t)\left[1+\delta y(t)-\tfrac12K_\theta(t,t)+\mathcal O(\delta y^2)\right], 
\end{equation}
reducing the calculation to the covariances of $\delta y(t)$, which obey $\mathrm{Var}[\delta y(t_i)]=K_\theta(t_i,t_i)$ and $\mathrm{Cov}[\delta y(t_i),\delta y(t_j)]=K_\theta(t_i,t_j)$. The first term is calculated exactly, with
\begin{equation}
\mathrm{Var}\!\left[\sum_{i=1}^N \log c(t_i)\right]
=
\sum_{i,j=1}^N K_\theta(t_i,t_j).
\end{equation}
Using the expansion of $c(t)$ given above, the covariance between the discrete sum and the time-integrated concentration is
\begin{equation}
\mathrm{Cov}\!\left(\sum_{i=1}^N \log c(t_i),\, r\int_0^T c(t)\,dt\right)
=
r\sum_{i=1}^N \int_0^T K_\theta(t_i,t)\,\bar c_\theta(t)\,dt,
\end{equation}
while fluctuations of the integrated concentration contribute
\begin{equation}
\mathrm{Var}\!\left(r\int_0^T c(t)\,dt\right)
=
r^2\int_0^T\!\!\int_0^T \bar c_\theta(t)\bar c_\theta(t')K_\theta(t,t')\,dt\,dt'.
\end{equation}
Collecting these contributions and retaining terms up to second order in the covariance
$K_\theta$, we arrive at
\begin{multline}
\label{eq: expanded log posterior3}
    \ell(\theta) \approx  \sum_{i=1}^{N} \left[\log \bar{c}_\theta(t_i) - \frac{1}{2} K_\theta(t_i, t_i) \right]  - r \int_0^T \bar{c}_\theta(t) \,d{t} \\ + \frac{1}{2} \left[\sum_{i,j=1}^{N} K_\theta(t_i, t_j) - 2 r \sum_{i=1}^N \int_0^{T}K_\theta(t_i, t) \bar{c}_\theta (t) \, d{t} + r^2 \int_0^T \int_0^T \bar{c}_\theta(t) \bar{c}_\theta(t') K_\theta(t,t') \, d{t} d{t'} \right].
\end{multline} 
The first cumulant, $\langle S_\theta \rangle$ depends only on the mean concentration profile. Because the multiplicative noise has been normalized such that $\langle c(t)\rangle = \bar{c}_\theta(t)$, the time-integrated term reduces directly to the integral of the mean profile. The logarithmic contributions from individual binding events acquire a correction proportional to the local variance $K_\theta (t_i, t_i)$, reflecting the asymmetry introduced by averaging the logarithm of a fluctuating quantity.  
Together, these terms quantify how temporal correlations in the concentration profile reduce the effective independence of binding events, thereby modifying the information available about $\theta$.

We now need to choose a specific kernel. For intrinsic stochasticity, where the (relative) standard deviation of fluctuations typically decreases with the molecular concentration, the kernel would need to depend on $\theta$. Extrinsic stochasticity is defined with relative fluctuations independent of the molecular concentration itself, and it is typically dominant for eukaryotic systems \cite{raser2005noise}. Here we focus on it, and thus choose the fluctuation kernel as independent of the inferred parameter, $K_\theta(t,t') = K(t,t').$ 
In this case, all $\theta$--dependence enters only through the mean profile $\bar c_\theta(t)$, while the statistics of fluctuations are fixed. 

In this case, we see that
\begin{align}
\mathrm{Var}[c(t)] &= \langle c^2(t) \rangle - \langle c(t) \rangle^2 =  \bar{c}_\theta^2(t) e^{-K(t,t)} \langle e^{2 \delta y}\rangle  -  \bar{c}_\theta^2(t) = \bar{c}_\theta^2(t) \left[e^{K(t,t)} - 1\right] \\ \sqrt{\mathrm{Var}[c(t)]} &\approx \bar c_\theta(t)\sqrt{K(t,t)}.
\end{align}
Thus, this assumption corresponds to concentration fluctuations being proportional to the mean concentration.
The kernel $K(t,t')$ controls the temporal correlations of  relative (dimensionless) fluctuations, i.e. fluctuations normalized by the mean concentration.

Now, differentiating the log likelihood function given in Eq.~\eqref{eq: expanded log posterior3} and taking the $\theta$-independent kernel gives
\begin{equation}
\frac{d\ell}{d\theta} \approx \sum_{i=1}^{N} \frac{1}{\bar{c}_\theta(t_i)} \dc{\bar{c}_\theta(t_i)}{\theta} - r\int_0^T \dc{\bar{c}_\theta}{\theta} d{t} + \left[r^2 \int_0^T \int_0^T \dc{\bar{c}_\theta(t)}{\theta} \bar{c}_\theta(t') K(t,t') \, d{t} d{t'} -r \sum_{i=1}^N \int_0^T K(t_i, t) \dc{\bar{c}_\theta(t)}{\theta} d{t} \right].
\end{equation}

The first two terms of this expression are the same as the $K=0$ case in the main text. Thus, the second two terms determine leading order  correction to the MAP estimator for $K \neq 0$. For a fixed $\theta$, the third term is deterministic, while the final term fluctuates for particular binding sequences $\{t_i\}$. Because $\langle \sum \delta(t-t_i) \rangle_{\{t_i\}}= r \bar{c}_{\bar{\theta}}(t)$ (where $\bar{\theta}$ is the true theta, $\hat \theta = \bar{\theta} + \delta\theta$), 
\begin{multline}
\mleft\langle r^2 \int_0^T \int_0^T \dc{\bar{c}_\theta(t)}{\theta} \bar{c}_\theta(t') K(t,t') \, d{t} d{t'} -r \sum_{i=1}^N \int_0^T K(t_i, t) \dc{\bar{c}_\theta(t)}{\theta} d{t} \mright\rangle_{\{t_i\}} \\= r^2  \int_0^T \int_0^T \dc{\bar{c}_\theta(t)}{\theta} (\bar{c}_\theta(t')  - \bar{c}_{\bar{\theta}})K(t,t') \, d{t} d{t'} = O( K T \delta \theta) = O(K \sqrt{N}),
\end{multline}
as long as $\int K(t, t') \dd{t'}$ is finite. Further, the fluctuations in the sum are
\begin{equation}
\sqrt{\mathrm{Var}\mleft[\sum_{i=1}^N \int_0^T K(t_i, t) \pdv{\bar{c}_\theta(t)}{\theta} \,\dd{t}\mright]} = O(K \sqrt{N}),
\end{equation}
as long as $\int_0^{\infty} K(t,t') \dd{t'}$ and $\partial \bar{c}_\theta / \partial \theta$ are both bounded.  

The first two terms in $\dd{\ell} / \dd{\theta}$, which are the same as for the $K=0$ estimator, are $O(N)$, while the correction terms for small $K$ are $O(K \sqrt{N}).$ Thus, for large $N$ and small $K$, $\hat{\theta}$ differs from the $K=0$ estimator only at $O(K/\sqrt{N})$.

Let's consider the simplest case in which the mean concentration $\bar{c}_\theta(t) = c_\infty$ is a constant. Furthermore we take a time-translation-invariant noise $K(t,t')=K(\left|t-t'\right|)$, with characteristic amplitude $\sigma^2 = K(0)$ and decay timescale $\tau = \sigma^{-2}\int_0^\infty K(t) d{t} \ll T$.  
In this limit, boundary effects are negligible and the integrals over the kernel reduce to simple functions of its time integral. 
\begin{align}
\label{eq: ML}
\frac{d\ell}{d c_\infty} &\approx \frac{N}{c_\infty} - r T + \left[r^2 \int_0^T \int_0^T c_\infty K(t,t')\, d{t} d{t'} -r \sum_{i=1}^N \int_0^T K(t_i, t)  d{t} \right], \\
0 &\approx\frac{N}{c_\infty} - r T + \left[r^2 c_\infty \sigma^2 \tau T -r N \sigma^2 \tau   \right],\\
0 &\approx  N - r T c_\infty - r N \sigma^2 \tau c_\infty + r^2 \sigma^2 \tau Tc_\infty^2,\\
0 &\approx (N - r T c_\infty) (1 - r \sigma^2 \tau c_\infty),\\
c_\infty &\approx \frac{N}{r T}.
\end{align}
The quadratic equation for $c_\infty$ also has a second, spurious root which diverges like $\sigma^{-2}$,  but this root corresponds to a local minimum of the likelihood and is therefore not physically relevant. 
Indeed, although the second-order cumulant expansion formally produces an additional stationary point with positive curvature, this feature lies outside the regime of validity of the expansion and does not correspond to a true extremum of the full likelihood.  
More precisely, our expression for $\ell (\theta)$ relies on truncating the cumulant expansion of $\log\langle e^{S}\rangle$ at second order,
\begin{equation}
\ell \approx \kappa_1(S) + \frac{1}{2}\kappa_2(S), 
\end{equation}
which is controlled only when higher cumulants of $S$ are parametrically small.  Higher cumulants have higher power of $c_\infty$, specifically $k_{2n} \sim c_\infty^
{2n} K(t,t')^{n}$. Hence, the second root $c_\infty\sim \sigma^{-2} \sim 1/K$ lies outside the regime where expansion can be trusted, and we discard it. 

The Eq.~\eqref{eq: ML} for the estimator shows that weak temporally correlated fluctuations in the concentration profile do not bias the estimator to the leading order. 

The second derivative, again assuming $K(t,t')$ is independent of $\theta$, is
\begin{multline}
\frac{d^2\ell}{d\theta^2} \approx \sum_{i=1}^{N} \frac{1}{\bar{c}_\theta(t_i)} \frac{\partial^2\bar{c}_\theta(t_i)}{\partial \theta^2} -\sum_{i=1}^{N} \frac{1}{\bar{c}_\theta^2(t_i)} \left( \frac{\partial\bar{c}_\theta(t_i)}{\partial \theta}\right)^2  - r\int_0^T \frac{\partial^2 \bar{c}_\theta}{\partial \theta^2} d{t}  \\ +\left[r^2 \int_0^T \int_0^T \dc{\bar{c}_\theta(t)}{\theta} \dc{\bar{c}_\theta(t')}{\theta} K(t,t')\, d{t} d{t'} + r^2 \int_0^T \int_0^T \frac{\partial^2 \bar{c}_\theta(t)}{\partial \theta^2} \bar{c}_\theta(t') K(t,t')\, d{t} d{t'} -r \sum_{i=1}^N \int_0^T K(t_i, t) \frac{\partial^2 \bar{c}_\theta(t)}{\partial \theta^2} d{t} \right].
\end{multline}

Now we use  $ \left\langle \sum \delta(t-t_i) \right\rangle_{\{t_i\}} = r \langle c(t) \rangle =  r \bar{c}_\theta(t) $, where $\langle .\rangle _{\{t_i\}}$ means ensemble averaging the sum of $\delta $ functions, averaging over different realizations of binding sequences. This gives
\begin{equation}
\left\langle \frac{d^2\ell}{d\theta^2} \right\rangle_{\{t_i\}} \approx - r \int_0^T\frac{1}{\bar{c}_\theta(t)}\left( \frac{\partial\bar{c}_\theta(t)}{\partial \theta}\right)^2  d{t} + r^2 \int_0^T \int_0^T \dc{\bar{c}_\theta(t)}{\theta} \dc{\bar{c}_\theta(t')}{\theta} K(t,t') \, d{t} \,d{t'}.
\end{equation}

For the particular case of constant concentrations, this gives
\begin{equation}
\left\langle \frac{d^2\ell}{d\theta^2} \right\rangle_{\{t_i\}} \approx - \frac{r T}{c_\infty}  + r^2 T \tau \sigma^2.
\end{equation}
Thus, to the leading order, the normalized error is
\begin{equation}
\frac{\uncert}{c_\infty^2 }\approx \frac{1}{N -c_\infty ^2 r^2 T \tau \sigma^2} \approx  \frac{1}{N (1 - c_\infty r \tau \sigma^2) }  \approx \frac{1}{N (1 - c_\infty r \int_0^\infty K(t)dt) } \approx
\frac{1}{N}\left ( 1 + c_\infty r \int_0^\infty K(t)d{t}\right).
\end{equation}
In this regime, small Gaussian stochastic fluctuations in the concentration around the mean renormalize the  prefactor of the error without modifying its scaling with the number of binding events.  
 If in the second term we plug $c_\infty = N/rT$, we get
\begin{equation}
    \frac{\uncert}{c_\infty ^2 } = \frac{1}{N} + \frac{1}{T}\int_0^\infty K(t)dt 
\end{equation}
Thus, although temporal fluctuations in the concentration introduce additional uncertainty, their effect is perturbative when the correlation time is short.

\end{document}